\documentstyle[emulateapj,psfig]{article}

\def\etal{{et al.}\thinspace}
\def\beb{\begin{thebibliography}{999}}
\def\eeb{\end{thebibliography}}
\def\bi{\bibitem[]{}}
\def\be{\begin{equation}}
\def\ee{\end{equation}}
\def\bea{\begin{eqnarray}}
\def\eea{\end{eqnarray}}
\def\o{\over}

\submitted{Submitted January 27 2000; accepted May 21 2000}

\begin{document}

\tighten
\singlespace

\lefthead{S. Majumdar and B. Nath}
\righthead{Cooling flows and SZ Effect}

\title{On cooling flows and the Sunyaev-Zel'dovich effect}

\author{Subhabrata Majumdar$^{1,2}$ }
\affil{$^1$Joint Astronomy Programme, Physics Department,
Indian Institute of Science, Bangalore 560012, India}
\affil{$^2$Indian Institute of Astrophysics, Koramangala, Bangalore 
560034 India}

\bigskip
\author{Biman B. Nath$^3$}
\affil{$^3$Raman Research Institute, Bangalore 560080, India}
\bigskip\bigskip

\begin{abstract}
We study the effect of cooling flows in galaxy clusters on the Sunyaev-Zel'dovich
(SZ) distortion and the possible cosmological implications. The SZ effect, along
with X-ray observations of clusters, is used to determine the Hubble constant,
$H_{\circ}$. Blank sky surveys of SZ effect are being planned to constrain the
geometry of the universe through cluster counts. It is also known that 
a significant fraction of clusters has cooling flows in them, which changes
the pressure profile of intracluster gas. Since the SZ decrement depends essentially
on the pressure profile, it is important to study possible changes in
the determination of cosmological parameters in the presence of a cooling
flow. We build several representative models of cooling flows and compare
the results with the corresponding case of gas in hydrostatic equilibrium.
We find that cooling flows can lead to an overestimation of the
Hubble constant. Specifically, we find that for realistic models of 
cooling flow with mass deposition (varying $\dot m$ with radius), there
is of the order $\sim 10 \%$ deviation in the estimated value of the Hubble
constant (from that for gas without a cooling flow) 
even after excluding $ \sim 80 \%$ of the cooling flow region
from the analysis.
We also discuss the
implications of using cluster counts from SZ observations to constrain other
cosmological parameters, in the presence of clusters with cooling flows.

\end{abstract}

\section{Introduction}

       Galaxy clusters are extensively observed in optical, X-ray and radio
bands. In the radio band, a cluster can be observed in the the Rayleigh-Jeans
side of the
cosmic microwave background spectrum, as a dip in the brightness temperature,
due to Sunyaev-Zel'dovich effect (SZ) (Sunyaev \& Zel'dovich 1972;
for a comprehensive review, see Birkinshaw 1999). 
The SZ distortion appears as a decrement for wavelengths $ \geq 1.44$ mm (frequencies
$\leq 218$ GHz) and as an increment for wavelengths $\leq 1.44$ mm.
SZ effect has the advantage that 
the SZ
intensity, unlike that of the X-ray, does not suffer from the $(1+z)^{-4}$ cosmological 
dimming. As discussed by numerous authors (Birkinshaw \& Hughes 1994;
Silverberg et.al. 1997), one can combine the X-ray
and radio observations for clusters to determine cosmological parameters.
This has been done in the recent years to determine the Hubble constant 
$H_{\circ}$
(Birkinshaw 1999). 
The SZ signal is, however, weak and difficult to detect. Recent high signal to noise
detection have been made over a wide range in wavelengths using single dish observations:
 at radio wavelengths (Herbig et al, 1995, Hughes \& Birkinshaw 1998), millimeter
wavelengths (Holzapfel et al. 1997, Pointecouteau et al. 1999) and submillimeter
wavelengths (Komatsu et al. 1999). Interferometric observations have also been carried out
to image the SZ effect (Jones et al. 1993, Saunders et al. 1999, Reese et al. 1999,
Grego et al. 2000). Other than estimating $H_\circ$ combining  SZ and X-ray data, 
SZ effect alone can also be 
used to determine the cosmological mass density
$\Omega_{\circ}$ of the universe (Bartlett \& Silk 1994,
Oukbir \& Blanchard 1992, 1997, Blanchard \& Bartlett 1998). However, these
procedures
generally assume the cluster gas is spherical, unclumped and isothermal. 
Almost all clusters, however, show departures from these simplistic
assumptions with some to a large extent. 

    Departures from these simple assumptions can lead to systematic
errors in the determination of the different cosmological parameters (Inagaki et al
1995),
especially, it was seen that non-isothermality of the cluster can lead to a
substantial error in values of the cosmological parameters. Temperature
structure in a cluster can be the result of the shape of the gravitational
potential (Navarro et al 1997, Makino et al 1998), or it can arise due to the 
fact that the initial
falling gas in the cluster potential is less shock heated than the later falling
gas (Evrard 1990). In fact hydrodynamical simulations of isolated clusters also
show a definite temperature structure and can introduce error in the value
of $H_{\circ}$ when compared to the traditional isothermal $\beta$-models
(Yoshikawa 1998). 
Roettiger et al (1997) have shown that cluster mergers can result in deviations from both
sphericity and isothermality.
 Observationally, the main handicap arises from the fact 
that the
thermal structure of clusters are hard 
to measure, and the temperatures generally
taken in analyses are the X-ray emission weighted temperature, usually measured over a 
few core
radii. So, in general an isothermal description of the cluster is taken (or
sometimes a phenomenological temperature model based on the Coma cluster: for example
see Eq 73 of Birkinshaw 1999).  

       In this paper, we study another important phenomenon that can 
substantially change the temperature structure, {\it viz.} a cooling flow.
Cooling flows in clusters of galaxies (for an introduction, see Fabian et al 
1984) is a well established
fact by now, and it is seen that around $60 -90\%$ of clusters exhibit cooling
flows in their core with $\approx 40\%$ of them having cooling flows of more
than $100 M_\odot ~ yr^{-1}$ (Markevitch et al 1998, Peres et al 1998, Allen et al 1999).
In the largest systems, the mass deposition rate can be as high as $1000 M_\odot ~ yr^{-1}$
(Allen 2000).
The idealised picture of a cooling flow is as following:
Initially when the cluster forms, the infalling gas is heated from gravitational
collapse. With time this gas cools slowly and a quasi hydrostatic state emerges.
However in the central region, where energy is lost due to radiation faster
than elsewhere,
an inward `cooling flow' initially arises due to the pressure gradient
(Fabian 1994). 
This can modify the
SZ decrement and act as a systematic source of error in the determination of the
cosmological parameters.

Schlickeiser
(1991) has shown that free-free emission from cold gas in the cooling
flow can actually lead to an apparent decrease of the SZ effect at the centre.
Since the central cooling flow region is generally very small, the
isothermal $\beta$-model of cluster gas can still be used for the majority of 
cluster
region even for cooling flow cluster, with the extra precaution of excluding 
the central X-ray spike from the X-ray fit, and a
corresponding change made in  the fitting of the SZ decrement. 
This is only
possible for nearby clusters, however with well resolved cluster cores. 

        Naively, the change in the central SZ decrement $y(0)$ can be seen 
as follows :
For a non cooling flow cluster, the central decrement is given by the line of
sight integral of the electron pressure through the cluster centre along the
full extent of the cluster. If the cluster has a maximum radius $r_{cl}$, 
then the central
SZ decrement at RJ wavelengths can be written as $y(0) = -4{{\sigma_T}\over{m_e c^2}}
\int_{0}^{r_{cl}} p_e dl $. For a cluster with a cooling flow, let us
suppose that the electron pressure $p_e$, drastically falls below a certain radius
$r_s$, which is typically well inside the core of the cluster. The resulting
central decrement is then  $y(0) \approx  -4{{\sigma_T}\over{m_e c^2}}
\int_{r_s}^{r_{cl}} p_e dl $. Depending on the distance of $r_s$ from the
cluster centre [$r_s \approx (0.1 $~to ~$0.3) r_{core}$],there wiil be a change in
the value of $y(0)$  by $\approx 5\% - 25\%$. 
However, this simplistic view may
not be true. This estimate assumes that the pressure profile remains a
$\beta$ profile outside the radius $r_s$.
The pressure profile, however, 
need not follow the $\beta$ profile once cooling flow starts and it can
deviate from it substantially even for radii much larger than $r_s$.
As a matter of fact, there can actually be an
increase in the pressure for a large region inside the cooling flow,
before a sudden drop inside $r_s$. Since the usual proecedure for estimating
the Hubble constant depends on fitting $\beta$ profiles to the SZ and X-ray
profiles, to estimate $r_{core}$, this change in the pressure profile
due to cooling flow can distort the estimation of $r_{core}$ and hence, the
value of $H_\circ$ in a non-trivial way. We study this effect in detail
in later sections.

       In this paper, we have investigated the problem 
of cooling flow
induced change in the temperature and density profile, its effect on the SZ effect, and
its subsequent effect on the determination of cosmological parameters. 
In  \S
2  we briefly review  the physics of Sunyaev Zel'dovich effect; \S 3 is devoted
to the physics of cooling flows and discussing the cooling flow solutions; 
in \S 4 we look at the effect of cooling flow
solutions on SZ effect, both on the determination of $H_{\circ}$ and $\Omega_{\circ}$; 
we conclude in \S 5 with a brief comment on how this
work differs from other work and the relevance of this paper. 
       For the SZ effect, our notation and approach mainly follows that
described in Barbosa et al(1996).

\section{Determining Hubble constant with Sunyaev-Zel'dovich effect}
\subsection{The Sunyaev-Zel'dovich Effect}
    The integral of the electron pressure along any  line-of-sight 
through the cluster determines the magnitude of
the distortion of the apparent brightness temperature of the 
cosmic microwave 
background (CMB) due to SZ effect. 
This is quantified in terms of the Compton $y$-parameter:
\begin{equation}
y=
\int {dl}{{k_BT_{e}}\over{m_{e}c^{2}}}n_{e}\sigma_{T},
\end{equation}
where $k_B$ is the Boltzman constant, $T_e$ is the gas temperature, $m_e$ is
the electron rest mass, $n_e$ is the electron number density, $c$ is the 
velocity of light and $\sigma_T$ is the Thomson scattering cross-section. 
This occurs through the inverse-Compton scattering, by the hot intracluster gas,
of the CMB photons propagating 
through the cluster medium, and the energy transfer in this interaction between
hot electrons and CMB photons resulting in a distortion to the CMB spectrum.
The SZ surface brightness at a position  $\theta$ 
of the cluster with respect to the mean CMB intensity
is given by
\begin{equation}
{\delta\textit{i}_\nu (\theta) = \textit{y}( \theta)\textit{j}_\nu(x), }
\end{equation}

\noindent 
\textit{x} is a dimensionless frequency parameter 
\begin{equation}
x = {{h\nu} \over {kT_o}},
\end{equation}
where $h$ is the Planck constant, $\nu$ is the observing frquency and $T_0$
is the CMB temperature at the present epoch: $T_0 \sim 2.73$K.
The function ${j}_\nu(x)$ describes the spectral shape of the effect
\begin{equation}
\textit{j}_\nu(x) =
{{2(kT_{\circ})^3} \over {(hc)^2}}
 {{x^{4} e^{x}} \over {\left(e^{x} -1\right)}^{2}}
\left[{ {x}\over{\tanh\left(x/2\right)}}-4\right].
\end{equation}
Since the total photon number is conserved in the inverse Compton scattering process, upscattering
of the photons, the spectral dependance gets an unique shape,
through a decrement in the
brightness temperature at lower frequencies while an increase is observed
at higher frequencies.

The Sunyaev-Zel'dovich effect provides a unique observing approach to traditional methods ---
which use X-ray temperature and X-ray luminosity. The X-ray studies have a major
disadvantage because of `cosmological dimming',  the 
surface brightness of distant X-ray sources falls off as $(1 + z)^{-4}$,
and for this reason, obtaining samples of clusters at
cosmological distances is challenging.
The SZ effect has the distinct advantage of being independent
of the distance to the cluster.  
The SZ flux density from a cluster will diminish 
with distance to the cluster
as the square of the angular-size distance; in contrast to X-ray
flux densities from clusters which diminish as the square of the luminosity
distance to the cluster.
If the observations resolve clusters, particularly at lower redshifts,
the observed sky SZ temperature distribution will be sensitive to the thermal electron
temperature structure within the clusters; once again this may be contrasted
with X-ray emission images of cluster gas distributions which are mainly 
sensitive to the gas density distribution.

\subsection{Determination of Hubble constant}
    The method for the determination of Hubble constant using Sunyaev-Zel'dovich
effect uses two observable quantities : 1) $\Delta T/T$ of the CMB due to SZ effect; 2)
the X-ray surface brightness $S_X$ of the cluster. These can be written as
\be
{{\Delta T_{SZ}}\over{T}}(r) = -2\int_{l_{min}}^{l_{max}}{{k_B T_e}\over{m_e c^2}}
\sigma_T n_e dl ,
\ee
\be
S_X(r) = {{1}\over{4\pi{(1+z)}^{4}}} \int_{l_{min}}^{l_{max}} {{dL_X}\over{dV}} dl ,
\ee
where $r$ is the distance to the line of sight from the cluster centre, $l_{max}$
and $l_{min}$ give the extension of the cluster along the line of sight, 
${{dL_X}\over{dV}}$ is the X-ray emissivity and $dl$ the line element along the
line of sight. 
    The X-ray emissivity in the frequency band $\nu = \nu_1$ to $\nu_2$ can 
be written as
\be
{{dL_X}\over{dV}} = {n_e}^2 \alpha (T_e;\nu_1, \nu_2, z) ,
\ee
where
\be
\alpha (T_e;\nu_1, \nu_2, z) = {{2}\over{1+X}}{\left [{{2\pi}\over{3m_e c^2}} \right ]}^{1/2}
    {{16{e}^6}\over{3\hbar m_e c^2}} A (T_e;\nu_1, \nu_2, z) ,
\ee
where
\begin{eqnarray}
&& A (T_e;\nu_1, \nu_2, z) = \int_{u_1 (1+z)}^{u_2 (1+z)} {(k_B T_e)}^{1/2}
e^{-u}\nonumber\\ 
&&[{Xg_{ff}(T_e,u,1)
+(1 - X)g_{ff}(T_e,u,2)}] du .
\end{eqnarray}
In the above equations we have assumed primordial abundance of hydrogen and
helium and have set $X=0.76$, $e$ is the electron charge, $\hbar = h/(2\pi)$,
 $u \equiv 2\pi\hbar\nu / {k_B t_e}$, and $g_{ff}(T_e,u,Z)$ is the
velocity averaged Gaunt factor for the ion of charge Ze (Kellog 1975). 
     Traditionally, to model the cluster gas distribution one takes the following density and
temperature profiles (Cavaliere \& Fusco-Femiano 1978)
\be
n_e(r) = n_{e0}\left[ 1 + {\left ({{r}\over{r_{core}}} \right )}^2 
\right ]^{-3\beta /2} , 
\ee
\be
T_e(r) = T_{iso} = constant,
\ee
where $n_{e0}$ is the central electron density and $r_{core}$ the core radius of the
cluster. The above expressions are used as an empirical fitting model, and the
parameter $`\beta'$ is regarded as the fitting parameter. The equation holds for
$0 < r < R_{cluster}$, where $R_{cluster}$ is the maximum `effective' extension
of the cluster. Conventionally, $R_{cluster} = \infty$, and then from equations
(5),(6),(10),(11), we get
\begin{eqnarray}
{{\Delta T_{SZ}}\over{T}}(\theta) &= &-{{2 \sqrt{\pi} \sigma_T k_B T_{iso}} 
\over
{m_e c^2}} n_{e0} r_{core}  
{{\Gamma (3\beta /2 -1/2)}\over{\Gamma (3\beta /2)}} \nonumber\\
&& \times {\left [ 1 + {({{d_A \theta}\over{r_{core}}})}^{2} \right ]}^
{1/2 - 3\beta /2} ,
\end{eqnarray}
\begin{eqnarray}
S_X (\theta) &=& {{\sqrt{\pi}}\over{4\pi {(1+z)}^4}} \alpha {n^2_{e0}} r_{core}
{{\Gamma (3\beta  -1/2)}\over{\Gamma (3\beta )}} \nonumber\\
&& \times {\left [ 1 + {({{d_A \theta}
\over{r_{core}}})}^{2} \right ]}^{1/2 - 3\beta } ,
\end{eqnarray}
where $\Gamma (x)$ is the gamma function. Since both the
central CMB decrement and the X-ray surface brightness are observed, one can 
then combine
equations (12) and (13) to estimate the core radius as
\bea
r_{c,est} &=& {{{\left [ {{\Delta T (\theta)}\over{T}}
\right ]}^{2}_{obs}}\over{{S_X (\theta)}_{obs}}}
{{\Gamma (3\beta  -1/2){\Gamma (3\beta /2)}^{2}}\over{{\Gamma (3\beta /2
-1/2)}^{2}\Gamma (3\beta )}} \\ \nonumber
& & {{m_e^2 c^4 \alpha}\over{16 {\pi}^{3/2}{(1+z)}^4 {\sigma_T k_B T^{2}_{e,fit}}}}
{\left [ 1 + {({{\theta}\over{\theta_{X,core}}})}^2 \right]}^{-1/2}  .
\eea
In the above equation $\theta_{X,core}$ is the angular core radius observed in
X-ray, and $T_{e,fit}$ is the X-ray flux averaged temperature (obtained from fitting
the observed X-ray spectrum to the theoretical spectrum expected from isothermal
case). This X-ray emission weighted temperature is given by 
\be
T_{iso} \equiv {{\int_{0}^{r_{vir}} T_e(r) \alpha {n_e}^2 r^2 dr}\over
          { \int_{0}^{r_{vir}} \alpha {n_e}^2 r^2 dr}}  .
\ee		  
		  
The point to be noted is that $r_{vir}$ is the virial radius of the cluster and
its choice depends on the observer. If the temperature has a 
spatial structure then the $T_{iso}$ inferred from such a procedure may 
give different
values depending on how much of the cluster 
is taken in making the above average.
It has been seen (Yoshikawa et al 1998), that this can lead to a substantial 
change in SZ effect inferred, and thus to the value $H_{\circ}$.	  

The angular diameter distance $d_A$ can be approximated for nearby ($z \ll 1$)
clusters as
\be
d_A = {{cz}\over{H_{\circ}}}\left [ 1 + {{2\Lambda - \Omega_{\circ} -6}
\over{4}}z + O(z^2) \right ] .
\ee
Thus finally we have got an estimate of $d_A(z)$ as $r_{c,est}/ \theta_{X,core}$.
Now, if from other observations we know the cosmological parameters $\Omega_{\circ}$
and $\Lambda$, then one can estimate the Hubble constant
    
	 As can be seen from the above equations, the value of $H_{\circ}$ depends
crucially on the many assumptions of isothermality and $\beta$-model density
distribution of the cluster. Cooling flow changes both of these and so it can
significantly affect the value of the Hubble constant.

\section{Cooling flows in clusters}

\subsection{Preliminaries}

     From X-ray spectra of clusters, it is known that the continuum emission
is thermal Bremsstrahlung in nature and originates from diffuse 
intracluster gas with
densities $10^{-2}-10^{-4} ~ cm^{-3}$ and temperatures around $10^7 -10^8$ K.
The gas is usually believed to be in hydrostatic equilibrium. If, however,
the density in the inner region is large enough, so that the cooling time
is less than the age of the cluster,
then there is a
`cooling flow' (Fabian et al 1984 and references therein). Of course there would 
be a flow only when the dynamical time is also shorter than the cooling
time scale ($t_{age} > t_{cool} > t_{dyn}$).

The basic equations of cooling flow are:
\bea
{du \over dr} &=& -{u \over \rho_b} {d \rho_b \over dr} - {2 u \over r}
-{\dot{\rho_b} \over \rho_b} ,\nonumber\\
u{{du}\over {dr}}+ {{1}\over {\rho_b}} {{d(\rho_b \theta)}\over {dr}}
&=&  -
{{GM_t(r)}\over {r^2}} ,\nonumber\\
u{{d}\over{dr}}\left ({{3\theta}\over{2}} \right ) - {{\theta u}\over{\rho_b}}
{{d\rho_b}\over{dr}} &=& {{\rho \Lambda (\theta)}\over{{(\mu m_p)}^2}} ,
\label{eq:flow}
\eea	
where $ \theta=2k_B T/ \mu m_p$, $\mu$ is the mean molecular weight,
and $m_p$ is the proton mass. For steady flows with constant mass
flux, $ \dot{\rho}=0 $. This implies
$u=\dot{m}/4 \pi \rho r^2$ for steady flows. (Note that in
cooling flows both $u$ and $\dot{m}$
are negative. The subscripts $b$ refers to baryons and $t$ refers to 
total i.e baryons + dark matter. However, we assume the baryonic contribution
to the total mass negligible w.r.t to the dark matter contribution.) 

$M(r)$ describes the distribution
of the total mass and depends on the details of 
dark matter density profiles
(see below).	   
$\Lambda (\theta)$ is the cooling function defined so that $n_e n_p
\Lambda (\theta)$ is the rate of cooling per unit volume. 
We use an analytical fit to the optically thin 
cooling function as given by Sarazin \& White (1987),
\bea
&& \bigl ( {\Lambda (\theta) \over 10^{-22} \, {\rm erg} {\rm cm}^3
{\rm s}^{-1}} \big )=4.7 \times \exp \bigl [ - \bigl ({T \over
3.5 \times 10^5 {\rm K}} \bigr ) ^{4.5} \bigr ] \nonumber\\
&+&0.313 \times T^{0.08} \exp \bigl [ - \bigl ({T \over
3.0 \times 10^6 {\rm K}} \bigr ) ^{4.4} \bigr ] \nonumber\\
&+&6.42 \times T^{-0.2} \exp \bigl [ - \bigl ({T \over
2.1 \times 10^7 {\rm K}} \bigr ) ^{4.0} \bigr ] \nonumber\\
&+&0.000439 \times T^{0.35} \,.
\eea
This fit is accurate to within $4$\% accuracy, for a plasma
with solar metallicity, within
$10^5 \le  T \le 10^8$ K. For $10^8 \le T < 10^9$ K,
it underestimates cooling
by a factor of order unity (compared to the exact cooling function
, as in, e.g., Schmutzler \& Tscharnuter, 1993)
, and therefore is a conservative
fit to use, as far as the effect of cooling is concerned.

For non-steady flows, we
adopt the formalism of White \& Sarazin (1987), where the mass deposition
rate, $\dot{\rho}$, is characterised by a `gas-loss efficiency' parameter $q$.
 One writes 
$\dot{\rho}=q (\rho / t_{cool})$ where $t_{cool}$ is the local isobaric cooling
rate ($t_{cool}=5 k_B T \mu m_p/\rho \Lambda$). It has been found that
$q \sim 3$ models can reproduce the observed variation of mass flux
($\dot{m} \propto r$) (Sarazin \& Graney 1991). 
Fabian (1994) has noted that these models of White \& Sarazin (1987)
 yield good approximations
to the emission weighted mean temperature and density profiles for
cooling flow clusters. We also note that Rizza et al (2000) have used the steady flow
models of White \& Sarazin (1987) to simulate cooling flows.

We first discuss cooling flows  with $\dot{m}=$constant.
With $q=0$,
one can eliminate the density from Eq.~\ref{eq:flow} to get two
differential equations:

\bea
{{du}\o {dr}} &=& {u \over \left [ r^2(5\theta - 3u^2) \right ]}
\left [ 3GM - 10r\theta + {{\dot{m}}\o{2\pi}} {{\Lambda
(\theta)}\o {{uM}^2}} \right ] \nonumber\\
{{d\theta}\o {dr}} &=& {2  \over \left [ r^2(5\theta - 3u^2) \right ]}
\left [ \theta ( 2u^2 r -GM) -( u^2 -\theta)
{{\dot{m}}\o {4\pi}} {{\Lambda (\theta)}\o {{uM}^2}} \right ] 
\label{eq:steady}
\eea
These equations have singularities at the sonic radius
$r_s$ where $5\theta_s=3u_s$. 
A necessary condition of singularity is that the numerators of 
Eq.~\ref{eq:steady}
vanish at the sonic radius. Therefore (Mathews \& Bregman 1978)

\be
r_s =(3/10\theta_s) \left [ GM(r) + {{\dot{m} \Lambda (\theta_s)}\o {10\pi
\theta_s M^2}} \right ]
\ee

We have used two different dark matter profiles for the cluster. The first
model ( Model A) has been discussed earlier in the literature in the
context of cooling flows in cluster (White \& Sarazin 1987;
 Wise \& Sarazin 1993) with a 
density profile,
\be
\rho_d=\left\{ \begin{array}{ll}
{\rho_o \over 1+(r/r_{core})^2} + {\rho _{o,g} \over 1+(r/r_{c,g})^2} 
& \mbox{if $r < 237$ kpc} \\
{\rho_o \over 1+(r/r_{core})^2}  & \mbox{if $r > 237$ kpc}
\end{array}
\right.
\ee
Here $\rho_o=1.8 \times 10^{-25}$ gm cm$^{-3}$ and $r_{core}=250$ kpc
describe the profile of the cluster mass, and $\rho _{o,g}
=4.1 \times 10^{-22}$ gm cm$^{-3}$ and $r_{c,g}=1.69$ kpc describe
the profile of the galaxy in the centre of the cluster.

Model B does not have the galaxy in the center, and so it is 
described simply by $\rho =\rho_0 / [1+(r/r_{core})^2]$.

With these assumptions, the solutions for steady cooling flows,
$\dot{m}=$ constant) are fully characterized by
(1) the inflow rate, $\dot{m}$, and (2) the temperature of the gas $T_s$
at the sonic radius $r_s$. Obviously, the cooling flow solutions are
only valid within the cooling radius $r_{cool}$ where $t_{cool}=t_{age}$.
We assume a value of $t_{age}=10$ Gyr for all models. We assume that
outside the cooling radius, gas obeys quasi-hydrostatic equilibrium
(Sarazin 1986). Although this means matching the cooling flow solutions with
nonzero $u$ to $u=0$ solutions outside, in reality the velocity of gas at
the cooling radius is very small (for a $\dot{m}=100$ M$_{\odot}$ yr$^{-1}$
with $\rho \sim 10^{-26}$ gm cm$^{-3}$, at $r=250$ kpc implies a velocity of 
$30$ km s$^{-1}$), which is close to the limit of turbulence in the
cluster gas (Jaffe 1980), and  
smaller than the sound
velocity ($\sim 1.5 \times 10^3 \, (T/10^8 \, K)^{1/2}$ km s$^{-1}$).
The velocity of the flow at the cooling radius is, 
therefore, for all practical
purposes,
sufficiently small to be matched to the
solution of hydrostatic equilibrium outside. (In this approach, we avoid the
time consuming search for the critical value of $\dot{m}$ for which the
flow solutions behave isothermally at $r \rightarrow \infty$ (see Sulkanen
\etal 1993).)

As in the usual assumptions for the interpretation of SZ effect, we assume
that the gas outside the cooling radius is isothermal, with a constant
temperature profile. The density, therefore, obeys $\rho \propto
[1 +(r/r_{core})^2]^{-3 \beta /2}$, where $\beta=\mu m_p \sigma ^2 /k_B T_{iso}$,
and $T_{iso}$ is the temperature of the gas at and outside the cooling radius.

For models with non-zero $q$ (Model C: has the same mass profiles as Model A), 
the solutions are characterized by
$T_s$ and the value of $\dot{m}$ at the cooling radius, $\dot{m}_{cool}$.
Since a fraction of mass drops out of the flow in this case, the inflow
velocity need not rise fast and so it is possible to find completely
subsonic solutions. 

\medskip
\subsection{Cooling flow solutions}

We numerically solved the flow equations  for
the parameters listed in Table 1. 
The density, temperature and pressure
profiles for three cases are presented
in Figures 1, 2 \& 3 . We mark the position of $r_{cool}$  in
each case, and we mark $r_s$ for the cases of transonic flows (when $\dot m=$
constant). Beyond $r_{cool}$ we match a hydrostatic solution, as
explained above, for the respective potentials. We also present, 
for comparison, the behaviour
if the solutions outside $r_{cool}$ are assumed to extend
inwards (that is, if no cooling flow is assumed).
We will postpone the discussion on the
effect of these profiles on the SZ decrement to a later section, and only
discuss the qualitative aspects of the solutions here.

The solution A1 is similar to that presented by Wise \& Sarazin (1993)
(their Figure 1; although they chose to characterise the solutions by
the temperature at $r_{cool}$ and not $T_s$ as we have done here).
It is also similar (qualitatively) to the solution for A262 presented
by Sulkanen \etal (1989). As the latter authors have noted, the effect
of having a galactic potential in the center is to have a flatter 
temperature profile for $r > r_s$, than in the case of no galactic potential.
This aspect is clearly seen while comparing our solutions with and
without galactic potentials in the center. Our calculations for the
case without the central galaxy are admittedly flawed in the very inner
regions where the gas density is larger than the dark matter density,
which results in an incorrect determination of the graviational potential
in the inner region. However, this happens only inside a region $\sim
25$ kpc from the centre, and should not influence our final results
by a large extent. 

A word of explanation for the pressure profiles is in order here.
Naively speaking, it would appear that the pressure profile inside the
cooling radius should have lower values than the corresponding case
of hydrostatic equilibrium. The fact that it is not always so has been
noted in the literature (e.g., Soker \& Sarazin 1988, Fig 1 of Sulkanen \etal 1989
). The reason for the pressure bump just outside of the
sonic radius is that the flow in this inner region is not pressure driven,
but rather by gravity (see also Soker \& Sarazin 1988). 
This is why the bump in the profile depends
on the presence and absence of the galaxy in the centre. And this profile
leads to the curious result that
the presence of cooling flow can lead to the overestimation
of the Hubble constant as discussed in the next section

The model with mass deposition (C1) is shown in Figure 3. The
local mass flux is found to be almost proportional to the radius,
consistent with various observations (Fabian, Nulsen \& Canizares 1984;
Thomas \etal 1987),
and, therefore, is probably a realistic model for cooling flow clusters.
In this case the temperature drops gradually all the way through, since
the velocity does not rise too fast. The deposited mass is assumed to
impart negligible pressure and the pressure refers only to that of
gas taking part in the flow.

\section{Determination of Hubble constant}

In this section we discuss the SZ and X-ray profiles of clusters with cooling
flows. We compare these with profiles from cluster having gas in hydrostatic 
equilibrium, and comment on the reliability of measuring Hubble
constant.  
 The effect of
cooling flow and the subsequent increased Bremsstrahlung emmison is
seen in the sudden increase in the X-ray flux in the innermost region of 
the cluster 
(Figure 5). The
signature of the cooling flow is seen in the form of the central spike in
the X-ray profile.  The X-ray profile is only affected slightly by
the drop in temperature and it is the dependence on the gas density that
holds . The temperature dependance becomes important only near the sonic
point. Outside $r_{cool}$, the X-ray profile is the same as that in
the hydrostatic cases.
	   
   The SZ distortion is proportional to the 
line of sight integral of the pressure, and the sudden increase of the gas
   density inside the cooling radius is moderated by the decrease of the gas
   temperature. As a result there is a gradual increase in the gas pressure. 
   Near the sonic point the temperature drops drastically by orders
   of magnitude, and results in sudden decrease in pressure. However, since
   this change in pressure becomes acute only within $\approx 5\%$ of the core
   radius,  it contributes negligibly to the line-of-sight integral of the gas
   pressure, and leads to an increase in the SZ distortion inside the cooling
   radius for all models considered (see Figure 4). 
   Like the X-ray
   profiles, the SZ profiles outside $r_{cool}$ is the same as that for the
   corresponding hydrosstatic cases.
   
   The SZ profiles have been calculated in the Rayleigh-Jeans limit ($x \ll 1$) where
   $j_{\nu}(x)$ of Eqn. 4 goes to $-2$. In general, however, the profiles should be
   calculated using Eqn 2. Our results below are independent of the 
   observational frequency, since the profiles at different frequencies have similar
   shapes,
   with the amplitude of the SZ distortion scaled either up or down.
   
       Once both profiles are known, one can determine the deviation
in the value
   of the Hubble constant using Eqns(14) \& (16). 
 The deviation from the idealistic case can be parametrised as

\be
 f_H \equiv {{r_{core, true}}\o {r_{core,est,fit}}} =
 {{H_{0,est}}\o{H_{0,true}}}
\ee
The above formula has
been used to determine the deviation of the estimated value of $H_{\circ}$ from
the
actual value, for models listed in Table-1. The effect of cooling flow on
the determination of the cosmological parameters are summarised in Table-2.

  To begin
   with, one has to get best fitted values for $r_{core}$ (or $\theta_c$) 
   and $\beta$ from different profiles. Since, the estimation of
   the Hubble constant depends on the determination of these
   parameters from the profiles, we look at this issue in more detail.
   We must keep in mind that the best fitted value of  $r_{core}$ (or $\theta_c$) 
   and $\beta$ depends on whether one decides to fit the X-ray or the
   SZ profiles, and the choice can lead to significant
   differences in the estimated value of $H_{\circ}$. One of the reasons 
   for the strong dependance on the nature of the profile can be the 
   non-isothermality of the cluster
   gas. Recent observations indicate that intracluster gas has a temperature
   structure, see Markevitch et al 1998.
   This is because the y-parameter depends on the integral over $T_e$, while
   emissivity of thermal 
   Bremsstrahlung depends on  $\sqrt{T_e}$. The
   dependance of the Gaunt factor on $T_e$ isindirect and weak.
   Yoshikawa et al 1998 have shown that gas temperature drop in the central regions (their
   Fig. 3), should increase both $r_{core}$ and $\beta$ fitted to $y(\theta)$, and to
   alesser extent to $S_X(\theta)$, as compared to those compared to $n_e(r)$. This
   discrepancy increases at higher redshifts.
    However, in their case, there is little change in the gas density
   profile. Clumpiness can also give rise
   to different fits, resulting in an overestimation of the Hubble constant
   (Inagaki etal 1995). 
   
   There are two other important points that have to be kept in
   mind while fitting the profiles. First, we must remember that we are trying
   to fit a cluster having a finite profile with the formulae (Eqns 12 \& 13)
   for isothermal $\beta$ profiles which is derived assuming the cluster to be 
   of infinite
   extent. This can, by itself, lead to an overestimation of $H_{\circ}$
   (Inagaki et al 1995). Thus to have a good
   fit one must choose a segment of the profile 
   such that, within that segment, 
   the profiles (SZ or X-ray) for a finite cluster  
   do not differ much from those of a {\it hypothetical} 
   cluster of infinite extent. We found that SZ and X-ray profiles of 
   clusters
   start differing from those of infinite size at radii greater
   than $1.5$ times the core radius. Hence, we have restricted our fitting to
   radii within $1.5 r_{core}$.
   
   Next, one must also be careful to exclude the region close to the sonic
   point, so that the X-ray spike is excluded from the fit. 
   Also, the central portion in the SZ profile should be avoided as its
   inclusion can give an {\it
   apparent} central decrement less than its neighbouring points (see
   Schlickeiser 1991).
   We have fitted the SZ and X-ray profiles varying the inner cutoff radius and
   the results for a representative solution for each class of model 
   are tabulated
   in Tables 3 \& 4. Thus, all fittings were done for profiles extending from
   $r=r_{min}$ to $r=1.5r_{core}$. 

    As can be seen from Table-2, cooling flows can lead to an overestimation
 of the Hubble constant. However, we must emphasise, that it may not be possible
 to {\it a priori} estimate the amount of bias introduced in the measurement of
 the Hubble constant due to cooling flows.
 There is no simple correlation
 between the amount of cooling (i.e $\dot{m}$) and the change in the estimated
 $H_{\circ}$ from the actual value.
 The total change depends not only on $\dot{m}$, but
 also on the position of cooling radius, sonic radius, temperature
 at the sonic point and the isothermal temperature characterising the
 hydrostatic cases, with which comparisons are made. Specifically,
 the fitted values of $r_{core}(\theta_c)$ and $\beta$ for cooling flow
 models differ from hydrostatic models according to shape of underlying
profiles, which is marked by two important features, firstly the central excess
of X-ray flux (or excess decrement of SZ flux), and secondly the the deviation
from the
smooth  hydrostatic profile inside $r_{cool}$, the amount of overestimation
mainly depends on these factors.
  For models with a central
galaxy potential, there is always an over-estimation of $H_{\circ}$, which is
greater than the models without the central galaxy.

For the realistic cases of models C1 and C2, where we have a variable
$\dot{m}$ with $r$ inside the cooling radius, the deviation of estimated
Hubble constant from its actual value is almost the same. They are also greater
than that of models A and B, having similar mass flow rates. This may be due to
the fact that the maximum deviation in pressure from the hydrostatic cases is
more in non steady cases, than in steady flows. Also
non-steady cases are marked by the absence of the sonic radii and the
subsequent drop in temperature.

   We note that
   although the different choice of 
   fitting may change the absolute determination of 
   cosmological parameters, the trend i.e deviation from the correct values
   remains more or less unaffected. It is interesting to note that for B-type
   model (C1), which include mass deposition in cooling flows, the deviations
   decrease as one excludes a greater part of the cooling flow region
   (Table 4\& 5). The other
   models show an increase instead. 
    Here, we remind ourselves that models with
   mass deposition i.e C-type models are more realistic (Fabian 1994). It is 
possible that the unsually high value of deviation (Table 4) and the
counter-intuitive trend of increasing deviation with decreasing portion of 
cooling flow region used for fitting (Table 4 \& 5), arise because of the
unrealistic modelling of cooling flows. If we take the model B1 as a
realistic one, then Table 4 \& 5 show that to obtain a value of the Hubble
constant within an accuracy of $\sim 10 \%$, one should have $r_{min} \sim 0.8
r_{cool}$. In most cases, $r_{cool} < r_{core}$ (Fabian \etal 1984). However, as
$r_{cool}$ cannot be determined without actually detecting a cooling flow in
a cluster, 
we suggest that {\it a significant portion of the profile within $r_{core}$ 
should be excluded as a precaution}.
 The SZ and the X-ray profiles for the
   different models are shown in Figures 4 \& 5.

\section{Discussions and Conclusion}

Our work on the effect of temperature structure of clusters and its effect on
SZ decrement differs from other previous work of this nature in following way:
this
work takes into account the change in density profile as well as the temperature
profile since both becomes important in the central region of the cluster. Also,
previous authors have looked at the issue of non-isothermality of a cluster at 
radii greater than the core radius of the cluster, whereas we look at
temperature change at regions inside the core radius. For them the density
profile can still be well approximated by a $\beta$ profile, whereas for cooling
flow solutions, density profile is vastly different. Further, they have
neglected radiative cooling in their work. We for the first time look at SZ effect in
presence of radiative cooling, by first solving the cooling flow equations for
reasonable and physical solutions.

In summary, we find that the presence of a cooling flow in a cluster
can lead to an overestimation of the Hubble constant determined from
the Sunyaev-Zel'dovich decrement. 
We have used different models of cooling flows, with and without mass
deposition, and found the deviation in the estimated value of the Hubble
constant in the case of a cooling flow from that of hydrostatic equilibrium. 
We have used the usual procedure of
fitting the SZ and X-ray profile with a $\beta$ profile
 to get an estimated value of $r_{core}$, and then compared with that
for the case of gas in hydrostatic equilibrium in order to estimate the deviation
in the Hubble constant. For the more realistic models with mass deposition
(varying $\dot m$ with radius), we found that the deviation decreases with
the exclusion of greater portions of the cooling flow region. Quantitatively,
we found that for the deviation to be less than $\sim 10 \%$, one should
exclude a portion of the profile upto $\sim 0.8 r_{cool}$. Since $r_{cool}$
is difficult to estimate without actually detecting a cooling flow, we have
suggested that a significant portion of the profile inside $r_{core}$
should be excluded, to be safe.

There can be another important implication of the effect of cooling
flows. With the upcoming satellite missions (MAP \& Planck), we
have come to the point where there are efforts to constrain $\Omega_0$
with surveys of blank SZ fields (Bartlett et al 1998; Bartlett 2000),
ultimately giving rise to SZ-selected catalog of clusters (Aghanim
et al 1997). This method relies on estimating the number of SZ
sources brighter than a given threshold flux (Barbosa et al 1996). 
The point to be noted
is that since these surveys are essentially flux limited in nature,
the validity of the analysis to determine $\Omega_0$ depends 
crucially on the {\it one-to-one} association of flux-limits
to corresponding mass limits of clusters. From our analysis
above, it seems that it may not be possible to associate a unique
cluster mass to a given SZ distortion, given the uncertainty
due to the presence of cooling flows. This might lead to
contamination in SZ cluster catalogs and the inference of $\Omega_0$.
Recently, attempts have been made to constrain $\Omega_{\circ}$ from variance measurement 
of brightness temperature in blank fields (Subrahmanyan et al
1998) and comparing them to simulated fields (Majumdar \& 
Subrahmanyan, 2000), of cumulative SZ distortions from a cosmolgical
distribution of clusters. These results may also be systematically
affected due to the presence of clusters having cooling flows.

The estimations made in this paper strictly applies to cases where the image of the SZ
effect is directly obtained by single dish observations. For interferometric observations, 
the interferometer samples the fourier transform of the sky brightness rather than the
direct image of the sky. The fourier conjugate variables to the right ascension and
declination form the $u-v$ plane in the fourier domain. Due to spatial filtering by an
interferometer, it is necessary that models be fitted directly to the data in the $u-v$
plane, rather that to the image after deconvolution. We do not forsee drastic change from
our inferences in such cases since the result mainly depends on the deviation of the SZ and
X-ray profile in case of a cooling flow from those in hydrostatic equilibrium. This,
however, should
be looked in greater detail in future. We also note that with the growing number of high
quality images of the SZ effect with interferometers, which have greater resolution than
single dish antennas, the shape parameters of the clusters can be directly determined from
the SZ dataset rather than from an X-ray image (Grego et al. 2000).

Finally, we would like to add, though the calculations presented in this 
paper were done using the dark matter profile (Eqn. 21), which is "commonly
used" for calculating cooling flow solutions, it is inconsistent 
with the dark matter 
profile (Navarro et al 1997)  found in numerical simulations.
(For a comparison of mass and gas distribution in clusters having cooling flows
with different dark matter profiles, see Waxman \& Miralda-Escude',
1995).
Moreover, we have neglected the self gravity of the gas.  Suto et al (1998)
have calculated the effect of including the self gravity of the gas in determining the
gas density profile. To make strong conclusions about the effect of cooling
flows in the determination of the Hubble constant, one should take  both 
the points mentioned above into account.

\acknowledgments
We are grateful to Joseph Silk for his comments on the manuscript.
We also thank the referee, Dr Asantha Cooray, for numerous suggestions which have helped us
in improving the paper. 
One of the authors (SM) would like to thank the Raman Research Institute,
Bangalore, for hospitality and for providing computational facilities. 
SM acknowledges help recieved from R.Sridharan on using 
IDL.
Finally, he would
also like to express his gratitude to Pijushpani Bhattacharjee for his constant
encouragement. 
  


\newpage
\bigskip

{\footnotesize
{\bf Table 1} -- Parameters for cooling flow solutions

\noindent
\begin{tabular}{|c||c||c||c||c||c||c||c|c|}
\hline

Solution & Mass Model & q &$\dot{m}(r_{cool})$
& $T_s$ & $r_s$ & $r_{cool}$ & $T_{iso}$\\
\hline
&&&(M$_{\odot}$ yr$^{-1}$) & (K) & (kpc) & (kpc)& (K)\\

\hline
A1 & Model A &0 & 100 & $6.5 \times 10^6$ & $0.688$
& $127.5$ & $1.2 \times 10^8$\\

\hline
A2 & Model A &0 & 200 & $6.5 \times 10^6$ & $0.462$
& $96.1$ & $7.7 \times 10^8$\\

\hline
A3 & Model A &0& 300 & $6.5 \times 10^6$ & $0.712$
& $132.2$ & $7.7 \times 10^8$\\

\hline
B1 & Model B &0& 100 & $4.0 \times 10^6$ & $0.688$
& $85.7$ & $1.14 \times 10^9$\\

\hline
B2 & Model B &0& 200 & $6.5 \times 10^6$ & $0.462$
& $89.6$ & $1.9 \times 10^9$\\

\hline
B3 & Model B &0& 300 & $6.5 \times 10^6$ & $0.712$
& $110.3$ & $1.9 \times 10^9$\\

\hline
C1 & Model A &3& 200 &  & & $111.6$ & $1.1 \times 10^8$\\

\hline
C2 & Model A &3& 300 &  & & $132.2$ & $1.1 \times 10^8$\\

\hline
\end{tabular}
}
\bigskip

\bigskip

{\footnotesize
{\bf Table 2} -- Effect on central decrement and $H_{\circ}$ for
$r_{min}=0.1r_{core}$

\noindent
\begin{tabular}{|c|c|c|}
\hline

Solution Type  &  ${{H_{est}}\over{H_{true}}}$ & $\Delta y_0$ \\
\hline & & (\% change) \\

\hline
A1& $1.91$ & $35$ \\
\hline
A2& $1.18$ & $11.5$ \\
\hline
A3& $ 2.6$ & $25$ \\
\hline
B1& $1.36$ & $12.0$\\
\hline
B2& $1.19$  & $9.0$ \\
\hline
B3& $1.13$ & $8.5$ \\
\hline
C1& $ 2.6$ & $14.0$ \\
\hline
C2& $2.7$  & $11.3$ \\
\hline

\end{tabular}
}\bigskip

\bigskip

{\footnotesize
{\bf Table 3} -- Fitting of SZ profile and deviation of $H_{\circ}$ for
 $\dot{m}=200 m_{\odot}/yr$

\noindent
\begin{tabular}{|c|c|c|c|}
\hline

Solution Type &
           $r_{min}=0.2r_{cool}$&$r_{min}=0.5r_{cool}$&$r_{min}=0.8r_{cool}$ \\
\hline
& $H/H_{true}$ & $H/H_{true}$ & $H/H_{true}$ \\
\hline
A2 & $1.57$ & $1.74$ & $2.18$ \\
\hline
B2 & $1.44$ & $1.58$ & $2.06$ \\
\hline
C1 & $2.20$ & $1.18$ & $1.07$ \\
\hline

\end{tabular}
}\bigskip

\bigskip

{\footnotesize
{\bf Table 4} -- Fitting of X-ray profile and deviation of $H_{\circ}$ for
 $\dot{m}=200 m_{\odot}/yr$

\noindent
\begin{tabular}{|c|c|c|c|c|}
\hline

Solution Type &
	   $r_{min}=0.5r_{cool}$&$r_{min}=0.8r_{cool}$&$r_{min}=0.9r_{cool}$
		  &$ r_{min}=0.95r_{cool}$\\
\hline
& $H/H_{true}$ & $H/H_{true}$ & $H/H_{true}$ & $H/H_{true}$\\
\hline
A2 & -- &$4.7$ & $2.7$ & $1.7$ \\
\hline
B2 & -- &$4.9$ & $2.24$ & $1.6$ \\
\hline
C1 & $1.69$ &$1.12$ & $1.02$ & $ \approx 1.0$ \\
\hline
\end{tabular}
}\bigskip

\newpage

\begin{figure}
\plotfiddle{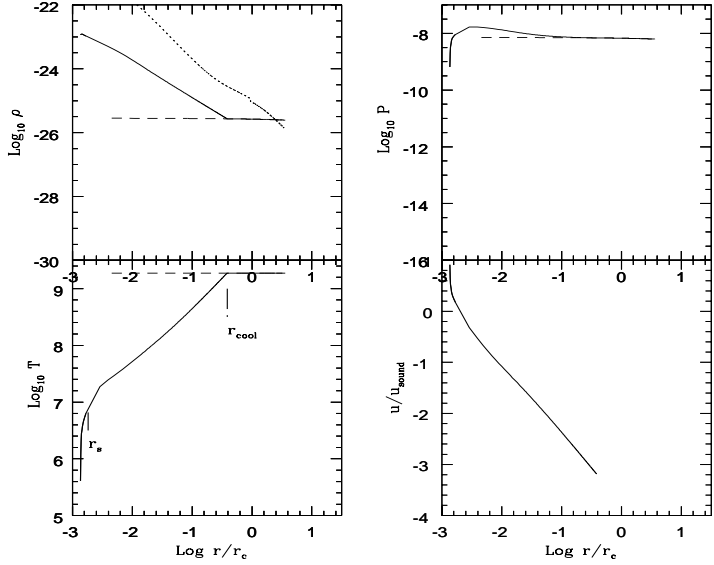}{3.0in} {0.0}{50.0}{40.0}{-225.0}{-50.0}
\caption {Cooling flow solution A2. The upper left panel shows the
dark  matter density profile (dotted line), the gas density profile
for the cooling flow (solid line) and the corresponding case of
gas in hydrostatic equilibrium. The lower left panel shows the
temperature profiles for the same cases. The position of $r_s$
and $r_{cool}$ are shown. The upper right panel shows the pressure
profiles and the lower right panel plots the Mach number of the
cooling flow gas as a function of the radius.}
\end{figure}

\begin{figure}
\plotfiddle{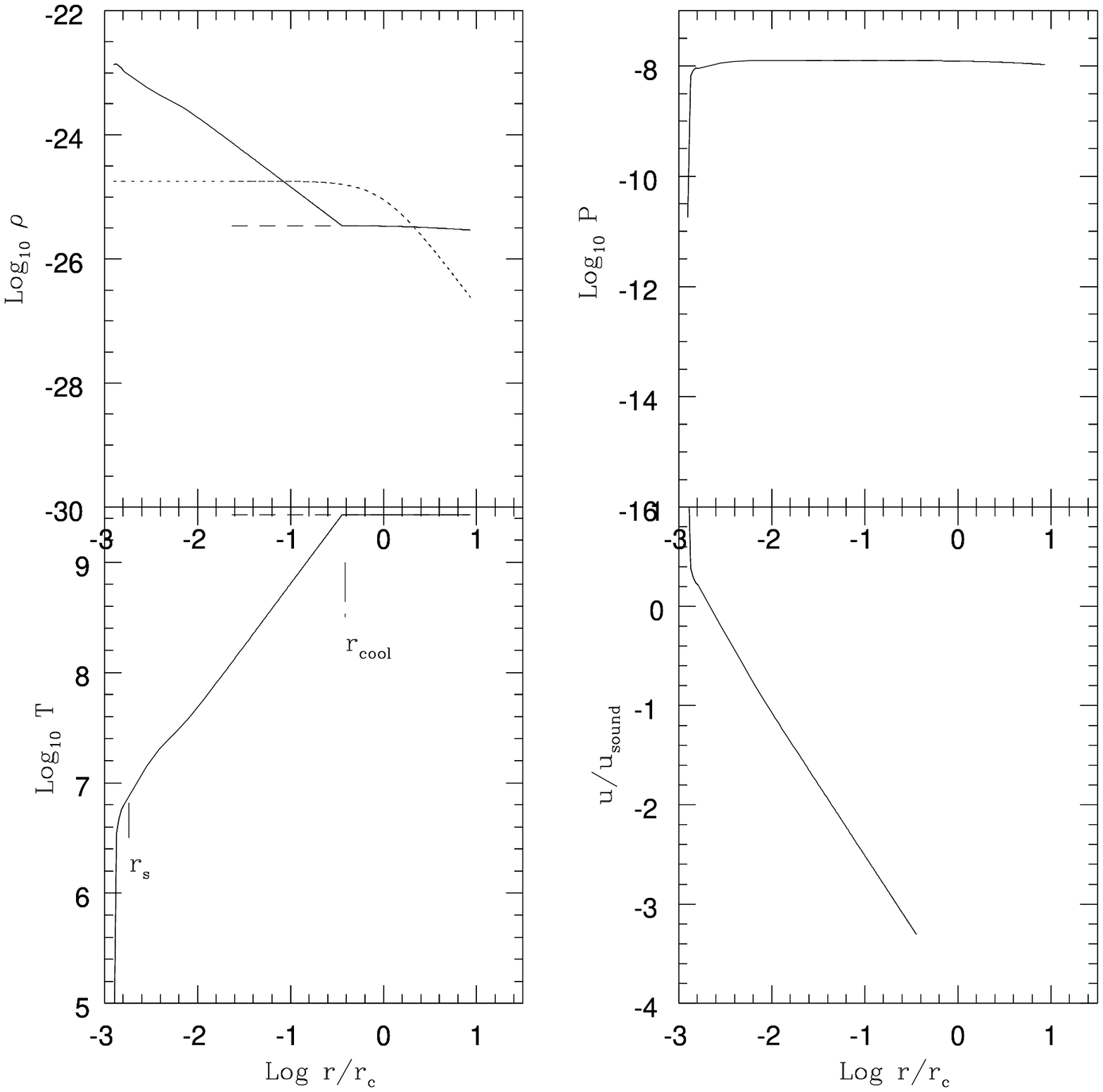}{3.0in} {0.0}{50.0}{40.0}{-225.0}{-50.0}
\caption {Same as figure 1, for cooling flow solution B2.}
\end{figure}

\begin{figure}
\plotfiddle{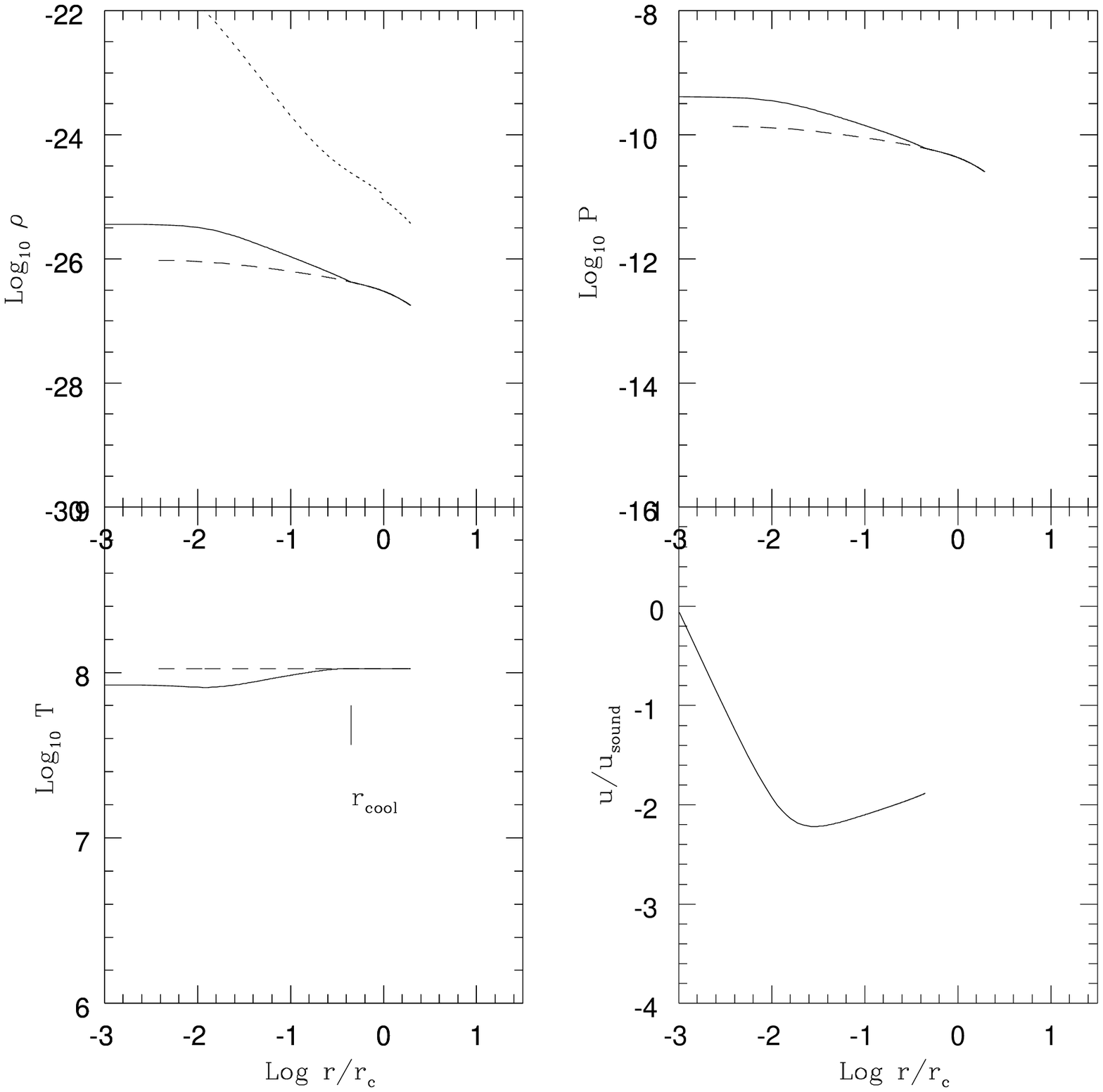}{3.0in} {0.0}{50.0}{40.0}{-225.0}{-50.0}
\caption {Same as figure 1, for cooling flow solution C1.}
\end{figure}

\begin{figure}
\centerline{
\psfig{file=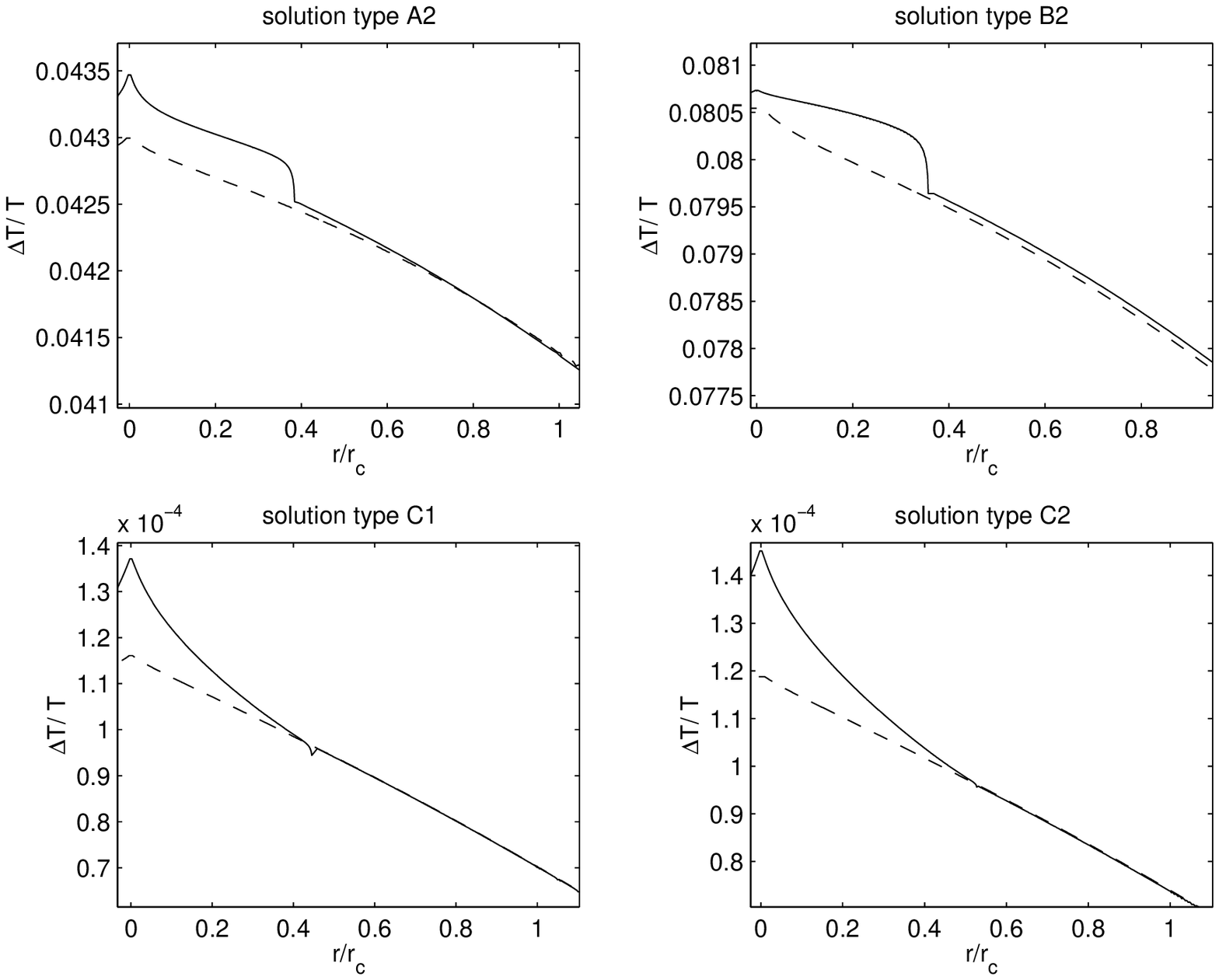,width=3.8in}}
\caption {Unconvolved SZ profiles,  
 for the solution types A2 (upper left),
B2 (upper right), C1 (lower left) and C2 (lower right), for cooling flow 
(solid line) and thecorresponding cases of gas in hydrostatic equilibrium 
(dashed line). ${{\Delta T_{RJ}}\over{T_{CMB}}}$ is plotted against ${{r}\over{r_{core}}}$.
Individual plots have been magnified to highlight the differences
between cooling flow and hydrostatic cases.} 
\end{figure}

\begin{figure}
\centerline{
\psfig{file=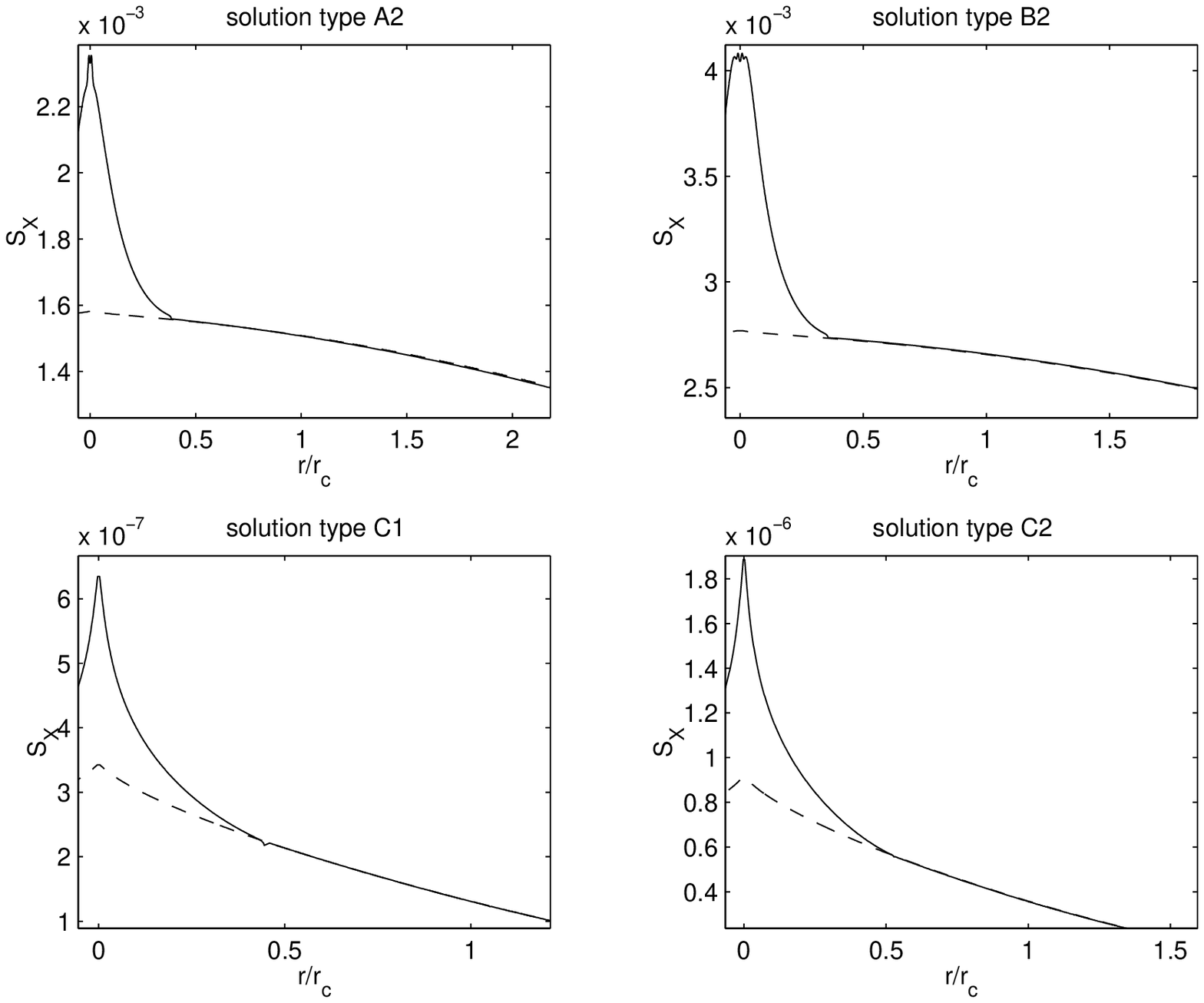,width=3.8in}}
\caption {Unconvolved X-ray profiles for the solution types A2 (upper left),
B2 (upper right), C1 (lower left) and C2 (lower right), for cooling flow 
(solid line) and the corresponding cases of gas in hydrostatic equilibrium 
(dashed line). X-ray surface brightness in units of
$erg~ s^{-1} cm^{-2} ster ^{-1}$ is plotted against ${{r}\over{r_{core}}}$.
Individual plots have been magnified to highlight the differences
between cooling flow and hydrostatic cases.} 
\end{figure}

\end{document}